\documentclass[12pt]{article}
\usepackage{amssymb}
\usepackage{amsmath}

\begin{document}

\title{\hspace{4.1in}{\small CERN-PH-TH/2004-179}\\
\hspace{4.1in}{\small OUTP 0420P}\bigskip \\
Wilson line breaking and gauge coupling unification.}
\author{ G. G. Ross$^{a,b}$ \\
$^{a}$ Theory Group, CERN, 1211 Geneva 23, Switzerland\\
$^{b}$Department of Physics, Theoretical Physics, University of Oxford,\\
1 Keble Road, Oxford OX1 3NP, U.K.}
\date{}
\maketitle

\begin{abstract}
We estimate the effect of threshold corrections coming from Wilson lines to
gauge coupling unification in the weakly coupled heterotic string with
orbifold compactification. By expressing the corrections in terms of an
effective field theory calculation we are able to estimate the minimal
threshold corrections in a realistic model without constructing the full
string theory. Using this we show that the effect of the gauge boson Kaluza
Klein excitations is systematically to reduce the string prediction for the
unification scale. In the case of discrete Wilson lines the effect of the
Wilson lines on both the gauge and matter sectors is fixed. We show that the
combined effect of the gauge and matter threshold corrections of the Kaluza
Klein excitations of the MSSM states can readily bring both the prediction
for the unification scale and for the strong coupling constant into good
agreement with experiment.
\end{abstract}

\section{Introduction}

The unification of gauge couplings \cite{georgiquinnweinberg} in
supersymmetric extensions of the Standard Model \cite{ibross+} remains the
most significant piece of quantitative evidence for physics beyond the
Standard Model. As discussed in \cite{precisiongr} the most accurate
prediction is obtained using the measured values of the strong and
electromagnetic couplings to determine the remaining gauge coupling. An
updated fit using the two-loop beta functions shows a consistent unification
of all three couplings at the unification scale $M_{X}=(2.5\pm 2)10^{16}GeV$
with a value of $\sin \theta _{W}$ given by $\sin ^{2}\theta _{W}=0.2334\pm
0.0025$ to be compared to the experimental value of $\sin ^{2}\theta
_{W}=0.2311\pm 0.0007\footnote{%
If instead one uses the measured values of $\sin ^{2}\theta _{W}$ and
electromagnetic coupling as input one finds the couplings unify at the scale 
$M_{X}=(4.3\pm 3)10^{16}GeV$ and $\alpha _{s}(M_{Z})=0.134\pm 0.005$ to be
compared to the experimental measurement $\alpha _{s}(M_{Z})=0.117\pm $ $%
0.017.$ This is consistent within the (rather poorer) theoretical error but
shows that a smaller value of $\alpha _{s}(M_{Z})$ would be better.}.$ The
error quoted in the unification predictions is dominated by the uncertainty
in the supersymmetric thresholds \cite{carena} - no error from possible
thresholds at the unification scale are included. The agreement between the
theoretical prediction and experiment is impressive, better than $1\%.$
Imposing the constraint that this precision should not be spoilt by the
effects of heavy states leads to strong restrictions on the underlying
theory \cite{precisiongr}.

Potentially equally important is the implication of the determination of the
unification scale because in some superstring theories this scale is
predicted in terms of the string scale or the Planck scale. This could
provide the first quantitative evidence for unification including gravity.
In particular in the weakly coupled heterotic string, without a stage of
Grand Unification below the compactification scale, the couplings are
predicted to unify at the scale $M_{s}$ which is close to the reduced Planck
scale \cite{kap,dienes} and is given by%
\begin{equation}
M_{s}=\frac{2\,e^{(1-\gamma _{E})/2}\,3^{-3/4}}{\sqrt{2\pi \alpha ^{\prime }}%
}\cong 0.527\,g_{s}\times 10^{18}\,GeV\simeq 3.6\times 10^{17}GeV  \label{ms}
\end{equation}%
This is only a factor of $8-15$ different from the central value of the
unification scale in the MSSM. Given the difficulty in determining the
unification scale some fifteen orders of magnitude above presently
accessible energies this is a remarkable result. However, in the context of
the weakly coupled heterotic string, it has proved to be difficult to bridge
the gap between these scales \cite{dienes,faraggi}. The addition of massive
matter states with appropriately chosen gauge quantum numbers can achieve it
but at the price of introducing significant sensitivity to the mass of these
new states and the related uncertainty to the prediction for $\sin \theta
_{W}$ \cite{precisiongr}. Another possibility is that there is a Grand
Unified group below the compactification scale. In this case one should
identify the gauge unification scale with the GUT breaking scale. However
this also introduces significant threshold sensitivity of $\sin \theta _{W}$
to the massive states at the GUT scale \cite{llswggr}. In addition one loses
the connection between the gauge unification scale and the string
unification scale.

The discrepancy between the gauge unification and weakly coupled heterotic
string scale has stimulated the search for alternative theories in which the
discrepancy might be resolved. In the strongly coupled heterotic string \cite%
{horwit} the string unification scale has an additional dependence on the
compactification scale of the eleventh dimension and agrees with the gauge
unification scale if the size of this dimension is of $O$($10^{15}GeV)^{-1}%
\cite{witten}.$ Of course, unless this compactification scale is known, one\
again loses the prediction for the gauge unification scale. There is some
indication that the $10^{15}GeV$ scale is special \cite{witten} but it is
not yet clear whether it is necessary for the compactification scale to be
at this value.

Subsequently there has been an explosion of interest in gauge unification at
much lower scales associated with low scale compactification \cite{dim}.
However, as argued in \cite{ghro}, the sensitivity to the new scales is such
that it is not possible to explain why the MSSM prediction for $\sin
^{2}\theta _{W}$ should be accurate to the 1\% level if there is a low
unification scale. Indeed the requirement that one should have minimal
sensitivity to heavy thresholds severely constrains the underlying string
theory to be one in which the compactification scale associated with gauge
nonsinglet states is close (within a factor of 2) to the string scale so
that most of the new massive states required by string unification lie\
above the string scale \cite{precisiongr}. In this case the sharp string
cutoff ensures the threshold effects of states more massive than the string
scale are small and in particular the power law sensitivity to thresholds
caused by the new space dimensions is avoided. If one further includes the
requirement of unification of the gauge interactions with gravity one even
obtains a constraint on the compactification scale of new dimensions in
which only gravity propagates given by $M_{c}>10^{15}GeV$.

Given these constraints on the ultra-violet completion of the theory it
seems that the weakly coupled heterotic string offers the best chance of
understanding why the MSSM gives such a precise prediction for $\sin \theta
_{W}.$ However this leaves the question of the discrepancy between the gauge
coupling scale and the unification scale unanswered. One possibility is that
threshold effects coming from Kaluza Klein\ and winding mode excitations are
responsible \cite{A15,nilles1}. However the initial estimates suggested that
these effects could close the gap only in the very large radius limit and
this spoils the precision prediction for $\sin \theta _{W}$ that we wish to
maintain. However one threshold effect has not been fully analysed, namely
the calculation of string threshold effects resulting from Wilson line
breaking. Although these have been calculated in specific string theories%
\cite{wl}, the analysis has not been done in realistic string theories.
However the indication coming from the toy models is that Wilson line
effects can be very significant even though the compactification scale,
which is related to the scale of Wilson line breaking for discrete Wilson
lines, is close to the Planck scale. As we have stressed this is a necessary
condition if the overall threshold sensitivity to unknown mass scales should
be small because, otherwise, one develops power law rather than logarithmic
sensitivity to heavy thresholds. In this paper we estimate the effect of
Wilson line breaking using the field theoretic calculation of the dominant
string effects \cite{ghilencea1} extended to include the effects of Wilson
line\ breaking \cite{ghilencea}. This allows us to discuss the minimal
corrections due to massive excitations of the states of the MSSM which are
present in all realistic compactified string theories. We can thus obtain a
quantitative estimate of Wilson line breaking effects for the case of an
orbifold compactification, without having an explicit construction of the
full string theory.

The result is encouraging. The Wilson line breaking does not shift the mass
of the Kaluza Klein (KK) excitations of the gauge bosons of the unbroken
gauge group. On the other hand the KK excitations of the broken generators
are split, half becoming lighter, half heavier, while the mean remains the
same. The\ contribution of the lighter states dominates and the net effect
(up to matter effects) is to reduce the string prediction for the
unification\ scale. This is what is required if the string unification scale
is to come closer to the gauge coupling unification scale. Moreover it turns
out that it is not necessary for the compactification scale to be made much
less than the string scale for these effects to be able to fill the factor
of $10$ gap. In this case power law running does not set in and the
sensitivity to the massive thresholds of the prediction for $\sin \theta
_{W} $ can still be below $1\%.$ Given that Wilson line breaking is very
often \textit{needed} to break the underlying gauge symmetry of the
heterotic string, this explanation seems very reasonable.

Of course it is necessary to include the effect of Wilson line breaking on
the matter states because these can spoil the MSSM predictions. It turns out
not to be possible to make general statements about their effects for
arbitrary Wilson lines, being sensitive to the magnitude and nature of the
Wilson line. In the case of discrete Wilson lines the magnitude is
determined and for simple discrete Wilson lines it is possible to determine
their effects on the matter states. Moreover, as we discuss, discrete Wilson
lines offer an elegant solution to the doublet triplet splitting problem 
\cite{wilson}. For these reasons we will concentrate here on the effects of
discrete Wilson lines.

To illustrate these effects we perform a quantitative study of the simple
case in which discrete Wilson lines associated with a $Z_{3}$ discrete group
break $SU(5)$ to the Standard Model. The effect is twofold. Firstly the mass
shift induced by the discrete Wilson line acts to increase the unification
scale taking it closer to the weakly coupled heterotic string prediction.
Moreover it can lead to a systematic reduction in the value for $\alpha _{s}$
which can eliminate the residual disagreement found in the MSSM.

\section{Wilson line breaking}

The Wilson line operator associated with the gauge group $G$ is defined as 
\begin{equation}
W_{i}=e^{i\int_{\gamma _{i}}dyA_{y_{m}}^{I}T^{I}},\quad I=1,\cdots ,rank\,G
\label{www}
\end{equation}%
where $A_{y_{m}}$ are the higher dimensional components of the gauge field, $%
y_{m}$ are the compact dimensions, $m=1$ ($m=1,2$) for one (two) compact
dimensions respectively. A summation over $m$ and $I$ is understood. $\gamma
_{i}$ are one-dimensional cycles of compactification. $T_{I}$ are the
generators of the Cartan sub-algebra of the Lie algebra

Continuous Wilson lines have their magnitude determined by a moduli field
which can have any vacuum expectation value. In this case they act just as
if the breaking was spontaneous and due to a Higgs scalar multiplet
transforming as the adjoint. The Wilson line breaks the associated gauge
group and gives a mass to those gauge bosons not commuting with the Wilson
line. Ignoring matter and the KK excitations, the mass of these excitations, 
$M_{X},$ defines the unification scale as above it there is no further
relative evolution of the gauge couplings of the unbroken gauge group
factors. Of course the overall gauge coupling continues to run until the
string cutoff scale but this effect can be absorbed in the starting value of
the gauge coupling used at $M_{X}.$ For matter fields the effect of Wilson
lines is determined by their gauge transformation properties. Chirality
protects those fields which are massless before Wilson line breaking from
acquiring a mass. However their KK modes can be shifted in mass by the
Wilson line and this can lead to further corrections to the gauge couplings.

If one is to discuss the effect of threshold effects close to the string
scale it is necessary to know the full spectrum of massive states. We will
be concerned with determining these effects for the minimal set of fields
consistent with obtaining the MSSM at low scales. In a theory such as the
heterotic string with an underlying GUT the main uncertainty in the spectrum
is the origin of the doublet triplet splitting needed if one is to have the
light Higgs doublets of the MSSM necessary for electroweak breaking. In the
case of continuous Wilson lines this must come from some mechanism, such as
the missing partner mechanism \cite{nanopoulos}, which significantly
complicates the minimal spectrum needed and gives rise to significant
uncertainties in the prediction for $M_{X}$ and $\sin ^{2}\theta $ following
from the associated threshold corrections. For this reason we concentrate on
the possibility that the Wilson lines are discrete because it provides a
natural solution to the doublet triplet splitting problem without
complicating the spectrum \cite{wilson}.

Discrete Wilson lines are associated with the projection of the states by a
discrete group, $D,$ which acts on the coordinates of the compactified $d-4$
dimensional space. In the absence of Wilson lines the physical states of the
theory correspond to discrete group singlet states. Thus the massless modes,
which necessarily have no dependence on the $4+d$ coordinates, must be
intrinsic $D$ singlets. When the discrete Wilson lines are switched on they
generate a representation, $\overline{D}\subset G,$ of the discrete group
which acts on the gauge quantum numbers of the gauge and matter states. In
this case the orbifold projection is modified so that the physical states of
the theory are singlets under the simultaneous action of the discrete group
action on both the internal and gauge quantum numbers. Thus if there is
massless representation, $R,$ of the gauge group, $G,$ that transforms as $S$
under the discrete group before Wilson line breaking then, with Wilson line
breaking, only the component of the representation that transforms as $%
S^{-1} $ under $\overline{D}$ remains, thus splitting the multiplet.

As a specific example consider the case that $G$ is $SU(5).$ In this case
the Higgs fields belong to the fundamental five dimensional representation $%
(R=5).$ Let us suppose this transforms as $S$ under the discrete group
before Wilson line breaking. Then if only the $SU(2)$ doublet component
transforms as $S^{-1}$ under the discrete Wilson line then only the doublet
remains light. The remaining colour triplet components are heavy, giving a
natural explanation for the doublet triplet splitting. In contrast to the
Higgs, the quark and lepton matter multiplets fill out complete $SU(5)$
representations. It is therefore necessary that these states should not be
split. This is also explained naturally in the case of discrete Wilson
lines.\ Division by a freely acting discrete group of order $N$ changes the
Euler number by the factor, $N.$ Thus in the theory before compactification
there must be an excess of $3N$ in the number of left-handed states
transforming as the $(\overline{5}\oplus 10)$ compared to the conjugate
representation. These states form complete $N$ dimensional representations
of the group $D$. As a result, after modding out by the diagonal subgroup of 
$(D\otimes $ $\overline{D}),$ one is left with $3$ families which form
complete representations of $\overline{D}$ and hence fill out complete $(%
\overline{5}\oplus 10)$ representations of $SU(5).$ However note that the
members of a family corresponding to different representations of $\overline{%
D}$ originate from different $D$ representations in the underlying manifold.

Note that in the discrete Wilson line case the massive gauge bosons,
associated with the broken generators, acquire mass quantised in fractions
of the compactification scale. The same is true of the heavy partners of the
Higgs doublets and the Kaluza Klein excitations. However our condition that
the precision prediction for gauge coupling unification should not be spoilt
by power law running requires that the compactification scale should be very
close to the string scale. Thus, in this case, it is the string scale that
determines the cutoff of the contribution of the zero modes and provides the
gauge unification scale. For a given compactified heterotic string model
with discrete Wilson line breaking no new parameters are introduced and
therefore the gauge unification scale remains a prediction.

In this paper we wish to consider (discrete) Wilson lines in orbifold
compactifications for the weakly coupled heterotic string. This means we
have to consider the KK and winding states that affect the running of the
gauge couplings. However in practice the fact that the gauge unification
scale is lower than the string scale means that the compactification radius, 
$R,$ is greater than the string length, $R_{string}=M_{s}.$ In this case the
KK states, whose mass is determined by integer multiples of $1/R,$ are less
massive than the winding states with mass integer multiples of $R$/$\alpha
^{\prime }.$ In the heterotic string contributions from states above the
string scale are exponentially \ suppressed and \ as a result the winding
mode, $m,$ has a contribution suppressed by a factor of $O(e^{-(2\pi
mRM_{s})^{2}}).$ Since this is small for the values of $R$ of relevance we
need consider the contribution of KK states alone. As we have stressed, in
order to calculate the effect of the Kaluza Klein modes on gauge unification
it is necessary to know their spectrum in detail. For this reason we must
consider specific compactification schemes and in this paper we consider
orbifold compactification in which the Kaluza Klein spectrum takes a
particularly simple form. As we shall see the dominant effect comes from
those Kaluza Klein states which, after Wilson line breaking, are anomalously
light. Although the precise details may change we expect the same effect to
occur in more general compactification schemes and so we think the results
we obtain in the context of orbifold compactification will be indicative of
the effects in more general Calabi Yau compactifications.

In supersymmetric compactifications the massive Kaluza Klein states belong
to $N=2$ or $N=4$ supermultiplets. Only the former contribute at one loop to
the relative evolution of the gauge couplings so we are principally
interested in these states. In orbifold compactification the $N=2$ sector is
associated with a $T^{2}$ torus subset of the original $T^{6}$. We will
consider the case of KK propagation two extra dimensions but first it is
instructive to consider the case of a single extra dimension.

\subsection{One compact dimension}

\subsubsection{One compact dimension}

In the case of one additional dimension the effect of Wilson line breaking
is to change the mass of the states in the KK tower according to 
\begin{equation}
M_{n}^{2}(\sigma )=\chi ^{2}+\frac{1}{R^{2}}(n+\rho _{\sigma })^{2},\qquad
\rho _{\sigma }=-R\!<\!A_{y}^{I}\!>\!\sigma _{I},\quad \text{(sum over $I$)}
\label{mass}
\end{equation}%
where $\rho _{\sigma }$ is derived using for one compact dimension with
constant $A_{y}^{I}$. $R$ is the compactification radius, $\sigma _{I}$ is
the weight or root of the representation that the higher dimensional field
(of charge $\sigma $ in Cartan-Weyl basis) belongs to. $\chi $ is the bare
mass of the KK tower that henceforth we take to be $0.$ The general
correction introduced to the gauge couplings by the combined effect of KK
states and Wilson lines is given by the general formula \cite{kap} valid
whether or not supersymmetry is present 
\begin{equation}
\Omega _{i}^{\ast }=\sum_{r}\sum_{\sigma =\lambda ,\alpha }\Omega
_{i}(\sigma ),\qquad \quad \Omega _{i}(\sigma )\equiv \sum_{n\in Z}^{\prime }%
\frac{\beta _{i}(\sigma )}{4\pi }\int_{0}^{\infty }\frac{dt}{t}\,e^{-\pi
\,t\,M_{n}^{2}(\sigma )/\mu ^{2}}\bigg\vert_{\text{reg.}}  \label{eq0}
\end{equation}%
$\Omega _{i}(\sigma )$ is the contribution to the one loop beta function for
the gauge coupling $g_{i}$ of a tower of KK modes associated with a state of
charge $\sigma $ in the Weyl-Cartan basis and of mass \textquotedblleft
shifted\textquotedblright\ by $\rho (\sigma )$ real. Here $\sigma =\lambda
,\alpha $ are the weights/roots belonging to the representation r. The sum
over $m$ runs over all non-zero integers and accounts for the effects of KK
modes of mass $M_{n}(\sigma )$ associated with the compact dimension. The
regulated sum over the KK tower can be performed and the gauge group
dependent piece is regularisation independent.

If the gauge symmetry group $G$ is a Grand Unified group before the Wilson
lines are \textquotedblleft turned on\textquotedblright , the overall
divergent part of $\Omega _{i}^{\ast }$ is the same for all group-factors $i$
that $G$ is broken into. As a result the $\sigma $ independent part of $%
\Omega _{i}^{\ast }$ can be absorbed into the redefinition of the tree level
coupling, similar to the case with one compact dimension. The resultant form
for the gauge couplings is \cite{ghilencea}%
\begin{equation}
\frac{1}{\alpha _{i}(R)}=\frac{1}{\alpha _{o}}-\sum_{r}\sum_{\sigma =\alpha
,\lambda }\frac{\beta _{i}(\sigma )}{4\pi }\ln \frac{\left\vert \,\sin \pi
\Delta \rho _{\sigma }\,\right\vert ^{2}}{\pi ^{2}\rho _{\sigma }^{2}}%
,\qquad \rho _{\sigma }=-R\,\sigma _{I}<\!A_{y}^{I}\!>\!  \label{sp1}
\end{equation}%
where $\rho =[\rho ]+\Delta _{\rho }$, $[\rho ]\in Z$ \ One only needs to
add here the contribution of zero modes\ (before Wilson line breaking), not
included in $\Omega _{i}^{\ast }$ and whose presence is in general model
dependent.

In the limit of turning off the Wilson lines vev's $\Delta _{\rho }=\rho
_{\sigma }\rightarrow 0$ the correction in (\ref{sp1}) coming from the KK
excitations vanishes and no splitting of the gauge couplings is generated.
The interesting case is what happens when $\rho _{\sigma }$ is non zero.

For the case of continuous Wilson lines the breaking can be continuously
taken to zero. The Wilson line acts in the same way as a physical Higgs
field providing the longitudinal component of the broken gauge bosons,
forming a massive $N=2$ supermultiplet. Thus for continuous Wilson lines
there is also a contribution from the ($m=0$) would-be zero mode which
acquires a mass $\rho _{\sigma }/R$ after Wilson line breaking. Including it
gives the result%
\begin{equation}
\frac{1}{\alpha _{i}(R)}=\frac{1}{\alpha _{o}}-\sum_{r}\sum_{\sigma =\alpha
,\lambda }\frac{\beta _{i}(\sigma )}{4\pi }\ln \frac{\left\vert \,\sin \pi
\Delta \rho _{\sigma }\,\right\vert ^{2}}{\pi ^{2}M_{s}R^{2}}.  \label{sp2}
\end{equation}

For the case that $\Delta \rho _{\sigma }$ is small this corresponds to a
contribution $\frac{\beta _{i}(\sigma )}{4\pi }\ln \left( \Delta \rho
_{\sigma }/M_{s}R\right) ^{2}$. The interpretation of this is
straightforward. As may be seen from eq(\ref{mass}) it corresponds to the
contribution of the KK state that has been made anomalously light, with mass 
$\Delta \rho /R$ through a cancellation of the Wilson line contribution to
the mass and the contribution associated with a non-zero KK level. The
string imposes a cutoff $M_{s}$ on the one loop contribution of this state
to the gauge coupling evolution. The contribution from all the higher levels
can be seen to be small corresponding also to the to the string cutoff
implicit in eq(\ref{eq0}).

\subsection{Two compact dimensions}

Assuming constant $A_{y_{1,2}}$, one computes the Wilson lines operator $%
W_{i}$ of eq.(\ref{www}) corresponding to each one-cycle $\gamma _{i}$ of
the two torus of compactification, $\rho _{i,\alpha }$ \cite{ghilencea}

\begin{equation}
\rho _{1,\alpha }=-R_{1}\alpha _{I}<\!A_{y_{1}}^{I}\!>,\qquad \rho
_{2,\alpha }=-R_{2}\,\alpha _{I}\Big[\!\!<\!A_{y_{1}}^{I}\!>\cos \theta
+<\!A_{y_{2}}^{I}\!>\sin \theta \Big]  \label{wilsonvev}
\end{equation}%
where $\theta $ is the angle between the two cycles and is fixed by the type
of orbifold considered ($\theta =2\pi /N$ for $T^{2}/Z_{N}$
compactifications). Using the Klein-Gordon equation in 6D the mass of the 4D
KK fields in the adjoint ($\alpha $) and fundamental ($\lambda $)
representations is given by \cite{ghilencea} 
\begin{equation}
M_{n_{1},n_{2}}^{2}(\sigma )=\frac{\mu ^{2}}{T_{2}\,U_{2}}\,|n_{2}+\rho
_{2,\sigma }-U(n_{1}+\rho _{1,\sigma })|^{2},\quad \sigma =\alpha ,\lambda .
\label{2d}
\end{equation}%
with the notation familiar in string models 
\begin{equation}
U\equiv U_{1}+iU_{2}=R_{2}/R_{1}\,e^{i\theta },\,\,\,\,(U_{2}>0);\,\qquad
T_{2}(\mu )=\mu ^{2}R_{1}R_{2}\sin \theta  \label{moduli}
\end{equation}%
%
%
%
%
%
%
%
%
%
%
%
%
%
%
%
%
%
%
%
%
%
%
%
%
%
%
%
%
%
%
%
%
%
%
%
%
%
%
%
%
%
%
%
%
%
%
For $\theta =\pi /2$ the two compact dimensions \textquotedblleft
decouple\textquotedblright\ from each other and in this case one finds $%
M_{n_{1},n_{2}}^{2}(\sigma )=(n_{1}+\rho _{1,\sigma
})^{2}/R_{1}^{2}+(n_{2}+\rho _{2,\sigma })^{2}/R_{2}^{2}$.

For generality we keep the $\theta $ angle arbitrary. If the gauge symmetry
group $G$ is a grand unified group before the Wilson lines are
\textquotedblleft turned on\textquotedblright , the overall divergent part
of $\Omega _{i}$ will be the same for all group-factors $i$ that $G$ is
broken into. As a result the $\sigma $ independent part of $\Omega
_{i}^{\ast }$ can be absorbed into the redefinition of the tree level
coupling, similar to the case with one compact dimension. In that case one
obtains the following splitting of the gauge couplings in 4D\cite{ghilencea} 
\begin{equation}
\!\!\frac{1}{\alpha _{i}}=\frac{1}{\alpha _{o}}+\sum_{r}\!\sum_{\sigma
=\alpha ,\lambda }\frac{\beta _{i}(\sigma )}{4\pi }\left\{ \ln \frac{\pi
e^{\gamma }|\rho _{2,\sigma }-U\rho _{1,\sigma }|^{2}}{(R_{2}\sin \theta
)^{2}}-\ln \bigg\vert\frac{\vartheta _{1}(\Delta _{\rho _{2,\sigma
}}-U\Delta _{\rho _{1,\sigma }}|U)}{\eta (U)}\bigg\vert^{2}\!\!\!+2\pi
\,U_{2}\Delta _{\rho _{1,\sigma }}^{2}\right\}  \label{ptcdr}
\end{equation}%
and with 
\begin{equation}
|\rho _{2,\sigma }-U\rho _{1,\sigma }|^{2}/(R_{2}\sin \theta )^{2}=|\sigma
_{I}\,(\!<\!A_{y_{2}}^{I}\!>\!-i\!<\!A_{y_{1}}^{I}\!>\!)|^{2}
\end{equation}%
As in the one dimensional case, the splitting of the gauge couplings is
induced by the combined effects of the compact dimensions and Wilson lines
vev's in a particular direction in the root space of the initial gauge group 
$G$. One may need to add to the above equation the correction from a zero
mode $(0,0)$ which acquires a mass after Wilson line breaking and the
massless states. Only a logarithmic correction will then be present with the
power-like corrections (divergences) \textquotedblleft
absorbed\textquotedblright\ into $1/\alpha _{o}$.

For the case that only $\rho _{2,\sigma }$ is non-zero the correction
approximately reduces to the form of eq(\ref{sp2}) after absorbing $\sigma $
independent terms in a redefinition of the coupling.

\subsection{Gauge Coupling Unification with Wilson line breaking}

In this Section we will use the results discussed above to determine the
effects of Wilson line breaking on gauge coupling unification. These
corrections have been determined in the context of the weakly coupled
heterotic string with orbifold compactification but the general structure is
indicative of the effects in more general compactifications because it is
driven by states which are made anomalously light by Wilson line breaking
and this happens in general compactification schemes.

The number of KK excitations contributing to the beta function is model
dependent, depending on how many (large) bulk dimensions the gauge field
propagates in and whether the KK modes fill out $N=2$ or $N=4$
representations. To determine the number we need to know the specific string
theory. However even without a specific theory we can estimate the magnitude
of the reduction in the string prediction for the unification scale to be
expected from Wilson line splitting of the KK excitations of the MSSM
fields. In this we are helped by the fact we are working in the limit where $%
R_{c}$ is larger than $R_{s}$. As discussed in \cite{raby}, in this limit
one can use either the full string theory or an effective field theory to
determine the KK spectrum. Thus we can use the intuition developed in
building orbifold GUTs when considering the minimal spectrum of KK
excitations.

\subsubsection{Kaluza Klein modes}

\paragraph{Kaluza Klein gauge excitations}

We start with a discussion of the effect of the gauge sector of KK modes.
The Wilson lines do not affect the gauge boson excitations associated with
the unbroken gauge group. The remaining $X$ and $Y$ gauge boson excitations
acquire mass corrections according to the form given in eq(\ref{2d}).

For the case of one additional dimension the contribution of the massive $X$
and $Y$ gauge bosons is given by eq(\ref{sp2}). On the other hand the
contribution of the KK modes of the unbroken gauge bosons is given by eq(\ref%
{sp1}) with $\rho _{_{3,2,1}}=0$ and the normalisation chosen is such that
this vanishes. The $X$ and $Y$ contribute to the relative gauge evolution in
the opposite way to that of the $3,2,1$ gauge bosons. From eq(\ref{sp2}) 
\footnote{%
Although we are considering discrete Wilson lines the equation still applies
because the $m=0$ field still contributes to the first massive level after
Wilson line breaking} we see they contribute between an \textquotedblleft
effective mass scale\textquotedblright\ given by $\frac{\left\vert \,\sin
\pi \Delta \rho _{\sigma }\,\right\vert ^{{}}}{\pi R^{{}}}$ and the cutoff
scale $M_{s}.$ Once they start to contribute the relative evolution of the
gauge couplings ceases corresponding (up to matter contributions) to a
reduction in the unification scale by the factor $\left\vert \sin \pi \Delta
\rho \,\right\vert /\pi RM_{s}$. Such a reduction is what is needed if the
unification scale in the weakly coupled heterotic string is to agree with
the gauge coupling unification scale. Note that this is a general conclusion
independent of the initial gauge group or the Wilson line. Given the
importance of this systematic trend it is of interest to see how it arises
directly from the form of the spectrum in eq(\ref{mass}) or eq(\ref{2d}).
From these equation it is clear that the effect of the Wilson line is
systematically to shift pairs of states with opposite signs of $n_{(1,2)}$
up and down in mass keeping the mean, $m,$ unchanged. In the field theory
calculation the logarithmic corrections coming from individual massive
states come in pairs with mass $m+\rho $ and $m-\rho $ giving the
contribution $\log (m+\rho )+\log (m-\rho )=\log (m^{2}-\rho ^{2}).$ We see
that the lighter state dominates and the net effect is a reduction from $%
m^{2}$ to $m^{2}-\rho ^{2}$ in the effective mass squared at which the
states start to contribute and systematically reducing the unification
scale. In the string theory calculation the string regularisation further
reduces the contribution of the more massive state going further in the
direction of reducing the unification scale.

In the calculation of the precise contribution of the massive states it is
necessary to compute their beta functions. As noted above, in the
calculation of the relative evolution of the gauge couplings, the
contribution of the $X$ and $Y$ gauge bosons is the same magnitude as the
contribution of the $SU(3)\otimes SU(2)\otimes U(1)$ Standard Model gauge
bosons but has the opposite sign. The contribution to the gauge coupling
running from massive KK modes comes only from the $N=2$ supermultiplets, the 
$N=4$ supermultiplets do not contribute at all. An $N=2$ gauge
supermultiplet contributes only $2/3$ of the contribution of an $N=1$ gauge
supermultiplet because it includes both an $N=1$ gauge \ supermultiplet and
an $N=1$ chiral supermultiplet.

\paragraph{Kaluza Klein matter excitations}

The minimal set of matter fields is that of the MSSM with 3 generations of
quarks and leptons and two Higgs doublets. Again the structure of their KK
excitations is model dependent depending on whether the propagate in the
bulk and if so in how many dimensions. In particular if they correspond to
twisted states about orbifold fixed points, they have no KK excitations. We
consider the various possibilities in turn.

\subparagraph{Twisted matter}

If all the matter fields correspond to twisted states then only the gauge KK
modes need be included. As discussed above the effect of the Wilson lines is
to reduce the scale at which the gauge contribution runs. However the
contribution of the matter fields is still cutoff at the string scale so
there is a mismatch between these scales. It is straightforward to determine
the net effect. At one loop order the quarks and leptons fill out complete
representations of $SU(5)$ and so do not change the relative evolution of
the gauge couplings which determine the unification scale and the precise
value of one of the three gauge couplings at low scales. For these
predictions, at this order, only the Higgs zero mode contribution and the
contribution of the gauge bosons need be included. If the Higgs contribution
were cut-off at the same reduced scale as the gauge bosons the prediction
would be just that in the MSSM with a reduced cut-off scale. However the
Higgs contribution is not cutoff and its contribution must be included
between the reduced unification scale and the original cut-off scale. This
changes the gauge coupling evolution by causing the electroweak coupling to
run more slowly. As a result, if the strong coupling is still to unify with
the other couplings, its value at low scales must be increased. In \ the
MSSM the value needed for the strong coupling is already somewhat larger
than the measured value so this change goes in the wrong direction. For this
reason we do not consider the twisted Higgs case further.

\subparagraph{Quark, lepton Kaluza Klein modes.}

If the quarks and leptons all have Kaluza Klein modes the situation is more
complicated as they all contribute to gauge coupling running. However for
the case of (discrete) Wilson lines their effect on the relative evolution
of the coupling constants vanishes at one loop because the massive quark and
lepton modes fill out complete multiplets of $SU(5)$ which are degenerate.
The reason is that these multiplets carry the same $(D\otimes $ $\overline{D}%
)$ intrinsic charge as is necessary if they are to give complete multiplets
of zero modes. As a result the massive excitations also have the same
dependence on the compactified coordinates and hence the same mass.

\subparagraph{Higgs Kaluza Klein modes}

If the Higgs fields are untwisted states they have KK excitations whose
effects need to be included. We consider the case that the doublet triplet
splitting is due to discrete Wilson line breaking. There are two cases to
consider.

If the Higgs doublet fields are discrete group, $D,$ singlets they must also
be $\overline{D}$ (Wilson line) singlets. This means that for them $\rho
_{\sigma }$ is zero in eq(\ref{wilsonvev}) and so their KK excitations are
unshifted. However this is not the case for their colour triplet partners
which are not $D$ singlets. At one loop order the contribution of the colour
triplet KK contribution to the relative gauge coupling evolution acts in the
opposite way to the Higgs doublets (the two together give no one loop
contribution). As a result the colour triplet states reduce the unwanted
increase in the strong coupling coming from the contribution of the light
Higgs above the gauge boson cutoff scale. The massive colour triplet states
belong to $N=2$ supermultiplets which means that they consist of two $N=1$
chiral multiplets and thus they have a beta function coefficient of
magnitude twice that of the Higgs (but opposite in sign). As a result they
can readily dominate over the \textquotedblleft excess\textquotedblright\
Higgs contribution above the reduced gauge boson unification scale and
actually reduce the value of the strong coupling, bringing it into better
agreement with experiment. We will present numerical estimates of this
effect below.

The other possibility is that the Higgs doublet fields come from non singlet 
$D$ fields. In this case the net effect of the KK doublet and colour triplet
fields is model dependent as both may contribute to the running of gauge
couplings.

\subsubsection{Wilson line breaking of $SU(5)$.}

Our discussion to date applies to a general Grand Unified gauge group, $G,$
before Wilson line breaking. However when making a quantitative estimate of
the effects we will illustrate the effects to be expected by considering the
simplest possibility with $G=$ $SU(5).$In this case the discrete group is
restricted to be $Z_{3}$\cite{distler}. The Wilson line group element is
given by $Diagonal(a^{2},a^{2},a^{2},a^{-3},a^{-3})\equiv e^{i\mathcal{Y}%
\theta }$ where $\mathcal{Y}=Diag[2,2,2,-3,-3].$ This breaks $SU(5)$ to $%
SU(3)\otimes SU(2)\otimes U(1)$ giving the remaining gauge bosons, $X$ and $%
Y $ gauge bosons a mass. The condition that this Wilson line should be a
representation of $Z_{3}$ is $\theta =2\pi n/3.$ In this case the Higgs
doublet fields are clearly $\overline{D}$ singlets which, as we discussed
above, is phenomenologically favoured.

For clarity of presentation we discuss the case of one additional dimension
but it is easy to generalise it to the case of two additional dimensions
using the results given above and the results in this case are included
below. To include the effects of the KK modes we use eq(\ref{sp1}). For our $%
SU(5)$ example with Wilson line breaking we have $\rho _{_{X,Y}}=5n/3.$ At
one loop order the massive modes affect the running of the couplings and
hence the unification predictions, giving 
\begin{eqnarray}
\alpha _{i}^{-1}(Q) &=&\alpha _{GUT}^{-1}+{\frac{b_{i}}{2\pi }}\log {\frac{Q%
}{M_{X}}}-\frac{2}{3}{\frac{b_{i}(3,2,1)}{4\pi }}\ln \frac{\left\vert \,\sin
\pi \Delta \rho _{24_{X,Y}}\,\right\vert ^{2}}{\pi ^{2}M_{s}^{2}R^{2}} 
\notag \\
&&+2\sum_{Higgs}\sum_{j}\frac{b_{i}(5_{j})}{4\pi }\ln \frac{\left\vert
\,\sin \pi \Delta \rho _{5_{j}}\,\right\vert ^{2}}{\pi ^{2}M_{s}^{2}R^{2}} \\
&\equiv &\alpha _{GUT}^{-1}+{\frac{b_{i}}{2\pi }}\log {\frac{Q}{M_{X}}+}%
\frac{b_{i}^{\prime }}{2\pi }  \label{coup}
\end{eqnarray}%
where the one loop beta function coefficient, $b_{i}$, is just that of the
MSSM, $b_{i}(3,2,1)$ is the coefficient coming from the unbroken (MSSM)
gauge supermultiplets, $b_{i}(5_{j})$ is the coefficient coming from the
Higgs supermultiplet sector\ and the sum $j=1,2$ is over the $(2,1)$ and $%
(1,3)$ $SU(2)\otimes SU(3)$ components of the $5.$ In this equation we have
absorbed terms independent of the group factor, $i$, in $\alpha _{GUT}^{-1}$
and we have $b_{i}=b_{i}(3,2,1)+b_{i}(5_{1}).$ The scale $M_{X}$ is the
unification scale taking account of the KK thresholds which is the value
that should be compared to the string prediction given in eq(\ref{ms}).$.$

We wish to determine the change in the unification scale and strong coupling
in this scheme relative to the MSSM in which 
\begin{equation}
\alpha _{MSSM,i}^{-1}(Q)=\alpha _{MSSM,GUT}^{-1}+{\frac{b_{i}}{2\pi }}\log {%
\frac{Q}{M_{X}^{0}}}  \label{coup1}
\end{equation}%
where $M_{X}^{0}$ is the unification scale in the MSSM. In both cases $%
\alpha _{1,2}(M_{Z})$ are input as their measured values. The unification
scale is found from the relative evolution of $\alpha _{1,2}$. Combining eqs(%
\ref{coup},\ref{coup1}), we find 
\begin{equation}
\log \left( {\frac{M_{X}}{M_{X}^{0}}}\right) ={\frac{b_{2}^{\prime
}-b_{1}^{\prime }}{b_{2}-b_{1}}}.  \label{newsc}
\end{equation}%
and 
\begin{equation}
\Delta \alpha _{3}^{-1}=-{\frac{1}{b_{1}^{\prime }-b_{2}^{\prime }}}\left[
(b_{2}^{\prime }-b_{3}^{\prime })\Delta b_{1}+(b_{3}^{\prime }-b_{1}^{\prime
})\Delta b_{2}+(b_{1}^{\prime }-b_{2}^{\prime })\Delta b_{3}\right] {\frac{1%
}{2\pi }}\log \left( {\frac{M_{X}^{0}}{M_{X}}}\right) ,  \label{nowed}
\end{equation}%
where $\Delta b_{i}$ are defined to be 
\begin{equation}
\Delta b_{i}=b_{i}^{\prime }-b_{i}.
\end{equation}

There are analogous expressions for the case $\alpha _{1,3}$ are input and
used to determine $M_{X}$ and $\sin ^{2}\theta _{W}.$

As discussed above the massive matter multiplets\ in a given representation
of $SU(5)$ are degenerate and so do not contribute to $\Delta b_{i}.$ Thus
in determining the changes in the unification scale and gauge coupling the
only contributions are the KK modes of the gauge bosons and the Higgs
multiplets.

In Table \ref{Table1} we give the results for the case $M_{s}R=2$. This the
largest value consistent with maintaining the precision prediction for gauge
couplings \cite{precisiongr} and large enough to justify our neglect of the
tower of winding modes \cite{string}. The numbers quoted apply to the case
of a $N=2$ massive spectrum in one extra dimension using the form of eq \ref%
{coup}.

\begin{table}[tbp] \centering%
\begin{tabular}{|l||l|l||l|l|}
\hline
& $n=1$ & $n=1$ & $n=2$ & $n=2$ \\ 
& $n_{H}=0$ & $n_{H}=2$ & $n_{H}=0$ & $n_{H}=2$ \\ \hline\hline
$\frac{M_{X}}{M_{X}^{0}}$ & 5.8 & 4.4 & 4.5 & 3.4 \\ 
$\Delta \alpha _{3}^{-1}$ & -0.33 & 0.47 & -0.29 & 0.52 \\ \hline\hline
$\frac{M_{X}}{M_{X}^{0}}$ & 4.7 & 6.0 & 3.7 & 4.8 \\ 
$\Delta \sin ^{2}\theta $ & 0.0011 & -0.0017 & 0.001 & -0.0018 \\ \hline
\end{tabular}%
\caption{\small The change in the string predictions for $M_sR=2$. The calculation refers to the 1D case, eq(\ref{sp1}), and is very close to the 2D case with $\rho_1 = \rho$ and $\rho_2 = 0$. The Wilson line group element, eq(\ref{mass}),  is specified 
by $a=e^{i 2\pi n /3}$. The columns with $n_H=0$ have no Higgs KK excitations while the columns with $n_H=2$ have KK excitations for both Higgs multiplets.   Finally the first two rows are obtained using $\alpha_{EM}$ and $sin^2\theta$ as input while the last two rows are obtained using $\alpha_{EM}$ and $\alpha_3$ as input.  }
\label{Table1} 
\end{table}%

From the Table one may see that, as expected, the unification scale is
increased bringing it closer to the string prediction. Without Higgs KK
modes there is an increase in the strong coupling which makes the agreement
with experiment worse. However with Higgs KK modes the strong coupling is
decreased improving the fit (a correction $\Delta \alpha _{3}^{-1}=1.1$
would brings the strong coupling into perfect agreement with experiment). If
instead we input the values of $\alpha _{1,3}$ the corrections to $\sin
^{2}\theta $ are small as expected and, with Higgs KK modes, brings the
result closer to the measured central value. For the case of Higgs KK modes
the unification scale is ($1.5\pm 2).10^{17}GeV$ if $\alpha _{s}$ and $%
\alpha _{em}$ are input or $(1.9\pm 2.5).10^{17}GeV$ if $\sin ^{2}\theta $
and $\alpha _{em}$ are specified. This is for $M_{s}R=2$ and the quoted
errors take into account the increased threshold sensitivity discussed
above. Overall one may see that the threshold effects have largely
eliminated the discrepancy between the observed and predicted unification
scales and they now agree within the errors associated with the residual
threshold uncertainties \footnote{%
If one increases $R$ it is possible to bring the string prediction right to
the \textquotedblleft experimental\textquotedblright\ value but as this
increases the residual threshold uncertainties well beyond the $1\%$ level
for $\sin ^{2}\theta $ we consider this unacceptable.}.

The result is sensitive to the form of the Wilson line. The reason is
because for $n=2$ the $m=0$ \textquotedblleft would-be\textquotedblright\
zero mode state is made heavier than the string scale and its contribution
is strongly suppressed. In this case we should use eq \ref{sp1} rather than
eq \ref{sp2} leading to the result shown in the second half of the Table.

It is very encouraging that the effect of the Wilson lines is to improve the
prediction for $\alpha _{s}$ and make the string prediction for the
unification scale closer to the \textquotedblleft
experimental\textquotedblright\ value. The latter provides the first
quantitative prediction of the string and if true would be evidence for an
underlying unification with gravity. However it is sensitive to the value of
the compactification radius, $R,$ and before we can treat the unification
scale as a prediction we must consider the range of values allowed for
viable compactifications scales. As we have stressed the minimum value for $%
R $ consistent with the precision prediction for $\sin \theta _{W}$
corresponds to $M_{s}R=2$ and Table \ref{Table1} has been calculated for
this value. The maximum value for $R$ is given by the self dual radius, $%
R_{sd},$ beyond which the duality transformation $R\rightarrow \alpha
^{\prime }/R$ takes one to a smaller value. This corresponds to $%
M_{s}R_{sd}=0.43.$ In fact the effect of the Wilson line breaking is
relatively insensitive to small changes in $R$. For $M_{s}R=1$ the scale
change is slightly reduced, lying between $3.4$ and $4.5.$ For smaller $R$
the field theory calculation is unreliable and winding mode contributions
may be significant. However even in this case we still expect the dominant
contribution to come from those modes below the string scale and these still
give a scale change of similar magnitude. Thus for the case of discrete
Wilson lines breaking $SU(5)$ with Higgs KK modes the indication is that the 
$M_{s}R=2$ estimate is representative and the string theory prediction for
the unification scale is in the range $(1.5\pm 2).10^{17}GeV.$

Although this calculation has been done in a field theory context with no
specific string compactification this contribution should be present in any
orbifold compactification and the general trend may be expected in more
general compactifications. For this reason the result is very encouraging
for the weakly coupled heterotic string. Of course it would be very
interesting to perform the full string theory calculation for a realistic
heterotic string compactification.

\section{Summary and conclusions}

The weakly coupled heterotic string has many attractive features as a
candidate for a theory of the fundamental interactions. The remarkable Grand
Unified aspects of the quark and lepton spectrum of the Standard Model are
readily explained because the heterotic string has an underlying GUT
structure. If supplemented with Wilson line breaking the string offers a
very elegant way of accommodating the Higgs doublets which do not fill out
complete GUT multiplets. Moreover in compactifications leading to a minimal
multiplet and gauge structure the scale at which the unification of gauge
couplings occurs is predicted.

In this letter we have estimated the massive KK threshold effects for the
minimal multiplet content on the gauge couplings of the massive states of
the theory after Wilson line breaking of the gauge symmetry. Remarkably we
find they systematically increase the unification scale bringing it close to
the weakly coupled heterotic string prediction and, if the Higgs states have
KK excitations, can reduce the prediction for the magnitude of the strong
coupling, bringing it into better agreement with experiment.

The result is sensitive to the compactification scale and for $M_{s}R$ as
small as $2$ the predictions for the ratio of gauge couplings and for the
weakly coupled heterotic string prediction of the unification scale are
close to the observed central values. For such small values of $M_{s}R$ the
power law running of the couplings does not develop between the
compactification scale and the string scale where the evolution of couplings
is cut off. As a result the sensitivity to the compactification threshold is
not great, allowing us to understand the remarkable accuracy of the
prediction for $sin^{2}\theta _{W}$ \cite{ghilencea} in the MSSM. These
results give quantitative support not only for a unification of the strong
weak and electromagnetic interactions but also for a unification of these
forces with gravity.

\bigskip

\textbf{Acknowledgements}

I\ am grateful to I. Antoniadis, P.Candelas, J.Distler, D.Ghilencea,
L.Ibanez, C.Lutken, J.March-Russell, K.Stelle for useful discussions and
particularly to F.Zwirner for his help and comments. This work was partly
funded by the PPARC rolling grant PPA/G/O/2002/00479 and the EU network
\textquotedblleft Quest for Unification\textquotedblright ,
MRTN-CT-2004-503369.

\end{document}